\documentclass[11pt]{article}
\usepackage{geometry}                
\usepackage{graphicx}
\usepackage{amssymb}
\usepackage{amsmath}
\usepackage{tikz}
\usepackage{float}
\usepackage{multirow}
\usepackage{epstopdf}
\usepackage{authblk}
\usepackage{booktabs}
\usepackage[all]{xy}
\begin{document}
\title{Cell Population Growth Kinetics in the Presence of Stochastic Heterogeneity of Cell Phenotype}
\author[1,2]{Yue Wang}
\author[3]{Joseph X. Zhou}
\author[3]{Edoardo Pedrini}
\author[3]{Irit Rubin}
\author[3]{May Khalil}
\author[4]{Roberto Taramelli}
\author[2]{Hong Qian}
\author[3,*]{Sui Huang}
\affil[1]{Department of Computational Medicine, University of California, Los Angeles, California, United States of America}
\affil[2]{Department of Applied Mathematics, University of Washington, Seattle, Washington, United States of America}
\affil[3]{Institute for Systems Biology, Seattle, Washington, United States of America}
\affil[4]{Department of Theoretical and Applied Science, University of Insubria, Italy}
\affil[*]{Corresponding author: sui.huang@isbscience.org}
\date{}                                           
\maketitle
\begin{abstract}
    Recent studies at individual cell resolution have revealed phenotypic heterogeneity in nominally clonal tumor cell populations. The heterogeneity affects cell growth behaviors, which can result in departure from the idealized uniform exponential growth of the cell population. Here we measured the stochastic time courses of growth of an ensemble of populations of HL60 leukemia cells in cultures, starting with distinct initial cell numbers to capture a departure from the {uniform exponential growth model for the initial growth (``take-off'')}. Despite being derived from the same cell clone, we observed significant variations in the early growth patterns of individual cultures with statistically significant differences in growth dynamics, which could be explained by the presence of inter-converting subpopulations with different growth rates, and which could last for many generations. Based on the hypothesis of existence of multiple subpopulations, we developed a branching process model that was consistent with the experimental observations. 
\end{abstract}

\smallskip
\noindent \textbf{Keywords.} 

\noindent leukemia; heterogeneity; branching process; growth pattern.

\section{Introduction}

Cancer has long been considered a genetic disease caused by oncogenic mutations in somatic cells that confer a proliferation advantage. According to the clonal evolution theory, accumulation of random genetic mutations produces cell clones with cancerous cell phenotype. Specifically, cells with the novel genotype(s) may display increased proliferative fitness and gradually out-grow the normal cells, break down tissue homeostasis and gain other cancer hallmarks \cite{hanahan}. In this view, a genetically distinct clone of cells dominates the cancer cell population and is presumed to be uniform in terms of the phenotype of individual cells within an isogenic clone. In this traditional paradigm, non-genetic phenotypic variation within one clone is not taken into account.

With the advent of systematic single-cell resolution analysis, however, non-genetic cell heterogeneity within clonal (cancer) cell populations is found to be universal \cite{pisco}. This feature led to the consideration of the possibility of biologically (qualitatively) distinct (meta)stable cell subpopulations due to gene expression noise, representing intra-clonal variability of features beyond the rapid random micro-fluctuations. Hence, transitions between the subpopulations, as well as heterotypic interactions among them may influence cell growth, migration, drug resistance, etc. \cite{tabassum,gunnarsson2020understanding,durrett2011intratumor}. Thus, an emerging view is that cancer, {even if we omit here the tumor tissue microenvironment}, is more akin to an evolving ecosystem \cite{Gatenby2014} in which cells form distinct subpopulations with persistent characteristic features that determine their mode of interaction, directly or indirectly via competition for resources \cite{Egeblad2010, Sonnenschein2000}. However, once non-genetic dynamics is considered, cell ``ecology'' differs fundamentally from the classic ecological system in macroscopic biology: the subpopulations can reversibly switch between each other whereas species in an ecosystem do not convert between each other \cite{Clark1991}. This affords cancer cell populations a remarkable heterogeneity, plasticity and evolvability, which may play important roles in their growth and in the development of resistance to treatment \cite{meacham_tumour_2013}. 

Many new questions arise following the hypothesis that phenotypic heterogeneity and transitions between phenotypes within one genetic clone are important factors in cancer. Can tumors arise, as theoretical considerations indicate, because of a state conversion (within one clone) to a phenotype capable of faster, more autonomous growth as opposed to acquisition of a new genetic mutation that confers such a selectable phenotype \cite{zhou2014multi,angelini2022model,howard2018multi,sahoo2021mechanistic,pisco,zhou2014nonequilibrium,kochanowski2021systematic}? Is the macroscopic, apparently sudden outgrowth of a tumor driven by a new fastest-growing clone (or subpopulation) taking off exponentially, or due to the cell population reaching a critical mass that permits positive feedback between its subpopulations that stimulates outgrowth, akin to a collectively autocatalytic set \cite{hordijk2018autocatalytic}? Should therapy target the fastest growing subpopulations, or target the interactions and interconversions of cancer cells? 

At the core of these deliberations is the fundamental question on the mode of tumor cell population growth that now must consider the influence of inherent phenotypic heterogeneity of cells and the non-genetic (hence potentially reversible) inter-conversion of cells between the phenotypes that manifest various growth behaviors and the interplay between these two modalities.

Traditionally, tumor growth has been macroscopically described as following an exponential growth law, motivated by the notion of uniform cell division rate for each cell, i.e., a first order growth kinetics \cite{Mackillop1990}. But departure from the exponential model has long been noted. To better fit experimental data, two major modifications have been developed, namely the Gompertz model and the West law model \cite{model}. While no one specific model can adequately describe any one tumor, each model highlights certain aspects of macroscopic tumor kinetics, mainly the maximum size and the change in growth rate at different stages. These models however are not specifically motivated by cellular heterogeneity. Assuming non-genetic heterogeneity with transitions between the cell states, the population behavior is influenced by many intrinsic and extrinsic factors that are both variable and unpredictable at the single-cell level. Thus, tumor growth cannot be adequately captured by a deterministic model, but a stochastic cell and population level kinetic model is more realistic.

Using stochastic processes in modeling cell growth via clonal expansion has a long history \cite{Zheng}. An early work is the Luria-Delbr{\"u}ck model, which assumes cells grow deterministically, with wildtype cells mutating and becoming (due to rare and quasi-irreversible mutations) cells with a different phenotype randomly \cite{Luria}. Since then, there have been many further developments that incorporate stochastic elements into the model, such as those proposed by Lea and Coulson \cite{Lea}, Koch \cite{Koch}, Moolgavkar and Luebeck \cite{Luebeck}, and Dewanji et al. \cite{dewanji}. We can find various stochastic processes: Poisson processes \cite{AS81}, Markov chains \cite{Gupta}, and branching processes \cite{Yue}, or even random sums of birth-death processes \cite{dewanji}, all playing key roles in the mathematical theories of cellular clonal growth and evolution. These models have been applied to clinical data on lung cancer \cite{lung}, breast cancer \cite{bre}, and treatment of cancer \cite{treat}.

At single-cell resolution, another cause for departure from the idealized uniform exponential growth is the presence of positive (growth promoting) cell-cell interactions (Allee effect) in the early phase of population growth, such that cell density plays a role in stimulating division, giving rise to the critical mass dynamics \cite{johnson2019cancer,korolev2014turning}.

To understand the intrinsic tumor growth behavior (change of tumor volume over time) it is therefore essential to study tumor cell populations in culture which affords detailed quantitative analysis of cell numbers or population size over time, unaffected by the tumor microenvironment and to identify departures from the idealized exponential growth. This paper focuses on stochastic growth of clonal but phenotypically heterogeneous HL60 leukemia cells at near single-cell sensitivity in the early phase of growth, that is, in sparse cultures. We and others have in the past years noted that at the level of single cells, each cell behaves akin to an individual, differently from another, which can be explained by the slow, correlated transcriptome-wide fluctuations of gene expression \cite{chang2008transcriptome,li2016dynamics}. Given the phenotypic heterogeneity and anticipated functional consequences, grouping of cells is necessary. Such classification would require molecular cell markers for said functional implication, but such markers are often difficult to determine a priori. Here, since most pertinent to cancer biology, we directly use a functional marker that is of central relevance for cancer: cell division, which maps into cell population growth potential –- in brief ``cell growth''.

Therefore, we monitored longitudinally the growth of cancer cell populations seeded at very small numbers of cells (1, 4, or 10 cells) in statistical ensembles of microcultures (wells). We found clear evidence that clonal HL60 leukemia cell populations contain subpopulations that exhibit diverse growth patterns. Based on statistical analysis, we propose the existence of three distinctive cell phenotypic states. We show that a branching process model captures the population growth kinetics of a population with distinct cell subpopulations. Our results suggest that the initial cell growth in the HL60 leukemic cells is predominantly driven by the fast-growing cell subpopulation. Reseeding experiments revealed that the fast-growing subpopulation could maintain its growth rate over several cell generations, even after placement in a new environment. Our observations underscore the need for treatment strategies that not only target the fast-growing cells but also the transition to them from the other cell subpopulations.

\section{Results}
\subsection{Experiment of cell population growth from distinct initial cell numbers.}
To expose the variability of growth kinetics as a function of initial cell density $N_0$ (``initial seed number''), HL60 cells were sorted into wells of a 384-well plate (0.084 $\mathrm{cm}^2$ area) to obtain ``statistical ensembles'' of replicate microcultures (wells) of the same condition, distinct only by $N_0$. Based on prior titration experiments to determine ranges of interest for $N_0$ and statistical power, for this experiment we plated 80 wells with $N_0=10$ cells ($N_0=10$-cell group), 80 wells with $N_0=4$ cells ($N_0=4$-cell group), and 80 wells with $N_0=1$ cell ($N_0=1$-cell group). Cells were grown in the same conditions for 23 days (for details of cell culture and sorting, see the Methods section). Digital images were taken every 24 hours for each well from Day 4 on, and the area occupied by cells in each well was determined using computational image analysis. We had previously determined that one area unit equals approximately 500 cells. This is consistent and readily measurable because the relatively rigid and uniformly spherical HL60 cells grow as a non-adherent ``packed'' monolayer at the bottom of the well. Note that we are interested in the \emph{initial} exponential growth phase and the apparent departure from the conventional exponential growth equation with uniform parameters that would apply to every replicate population. Here we are not studying in the latter phases when the culture becomes saturated and has historically been the focus of analysis (see Introduction).

\begin{figure}[ht]
	\centering
	\includegraphics[width=0.9\linewidth]{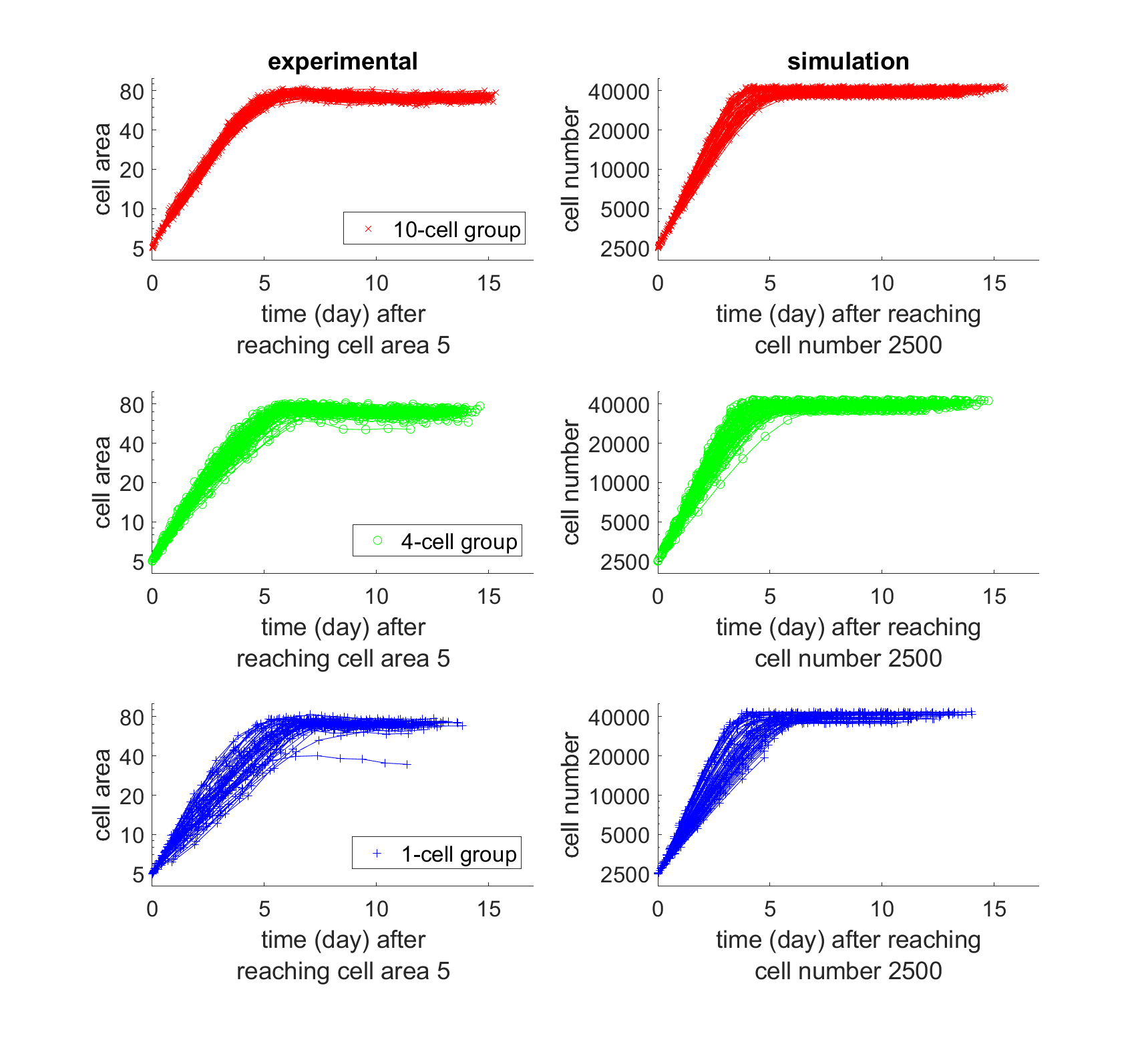}  
	\caption{Growth curves of the experiment (left) and simulation (right), starting from the time of reaching 5 area units (experiment) or having 2500 cells (simulation), with a logarithm scale for the $y$-axis. The time required for reaching 5 area units was determined by exponential extrapolation, as reliable imaging started at $>5$ area units. The $x$-axis is the time from reaching 5 area units (experiment) or 2500 cells (simulation). Red, green, or blue curves correspond to 10, 4, or 1 initial cell(s). Although starting from the same population level, patterns are different for distinct initial cell numbers. The $N_0=1$-cell group has higher diversity.
	}
	\label{f1}
\end{figure}

\begin{figure}[ht]
	\centering
	\includegraphics[width=0.9\linewidth]{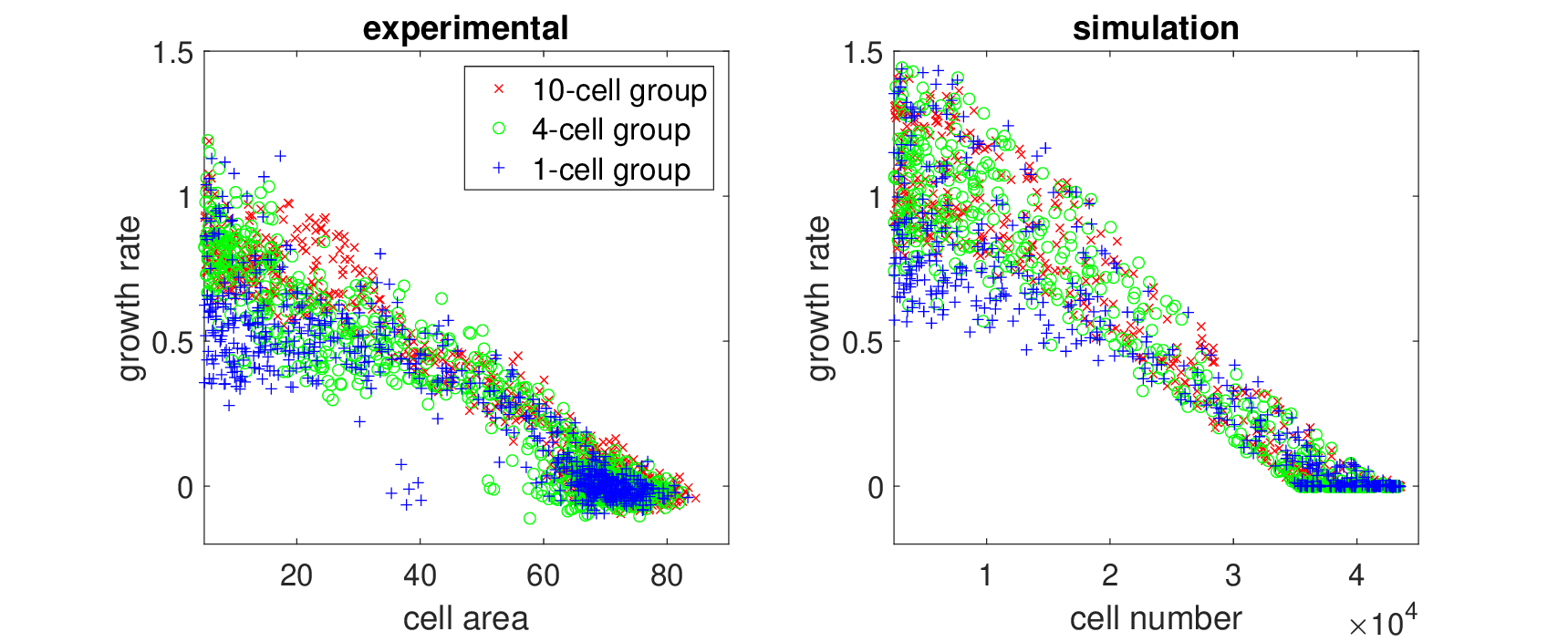}  
	\caption{Per capita growth rate (averaged within one day) vs. cell population for the experiment (left) and simulation (right). Each point represents one well in one day. Red, green, or blue points correspond to 10, 4, or 1 initial cell(s).}
	\label{f2}
\end{figure}

Wells that have reached at least 5 area units were considered for the characterization of early phase (before plateau) growth kinetics by plotting the areas in logarithmic scale as a function of time (Fig.~\ref{f1}). All the $N_0=10$-cell wells required 3.6-4.6 days to grow from 5 area units to 50 area units (mean=4.05, standard deviation=0.23). For the $N_0=1$-cell wells, we observed a diversity of behaviors. While some of the cultures only took 3.5-5 days to grow from 5 area units to 50 area units, others needed 6-7.2 days (mean=5.02, standard deviation=0.75). The $N_0=4$-cell wells had a mean=4.50 days and standard deviation=0.44 to reach that same population size. 

To examine the exponential growth model, in Fig.~\ref{f2} (left panel) we plotted the per capita growth rate against cell population size, where each point represents a well (population) at a time point. As expected, as the population became crowded, the growth rate decreased toward zero. But in the earlier phase, many populations in the $N_0=1$-cell group had a lower per capita growth rate than those in the $N_0=10$-cell group, even at the same population size -- thus departing from the expected behavior of {a uniform exponential growth model that describes cultures that nominally vary only $N_0$}. Analysis of variance (ANOVA) and weighted Welch's $t$-test showed that the difference in these growth rates was significant (see the Methods section).

While qualitative differences in the behaviors of cultures with different initial seeding cell numbers $N_0$ can be expected for biological reasons (see below), in the elementary exponential growth model, the difference of growth rate should disappear when populations with distinct seeding numbers are aligned for the same population size that they have reached as in Fig.~\ref{f2}. 

A simple possibility is that the deviations of expected growth rates emanate from difference in cell-intrinsic properties. Some cells grew faster, with a per capita growth rate of $0.6\sim 0.9$ (all $N_0=10$-cell wells and some $N_0=1$-cell wells), while some cells grew slower, with a per capita growth rate of $0.3\sim 0.5$ (some of the $N_0=1$-cell wells). In other words, there was intrinsic heterogeneity in the cell population that is not ``averaged out'' in the culture with low $N_0$, and the sampling process exposes these differences between the cells that appear to be relatively stable.

\begin{figure}[ht]
	\centering
	\includegraphics[width=0.9\linewidth]{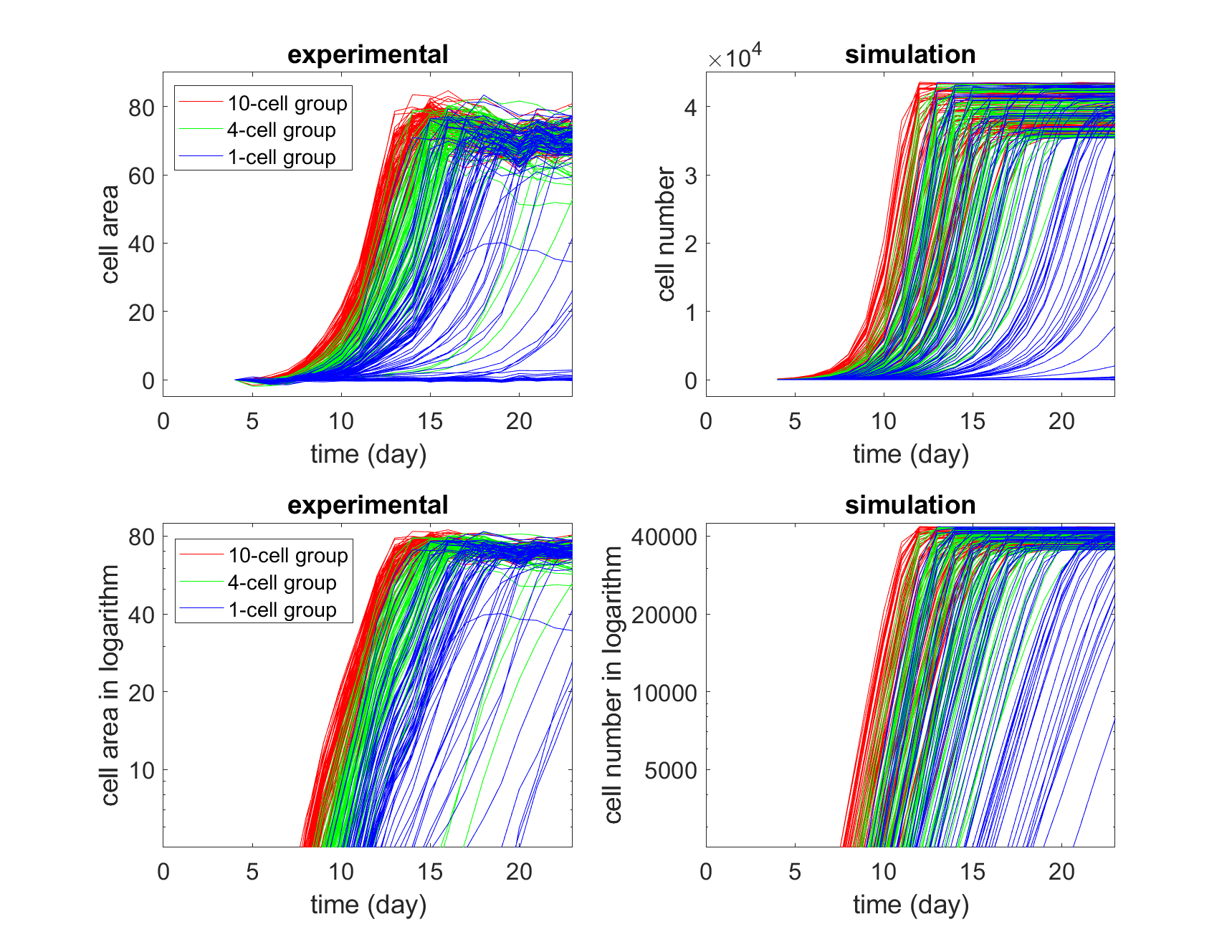}  
	\caption{Growth curves of the experiments with different initial cell numbers $N_0$ (left) and growth curves of corresponding simulation (right). Each curve describes the change in the cell population (measured by area or number) over a well along time. Red, green, or blue curves correspond to $N_0=10$, $4$, or $1$ initial cell(s). {For the upper panel, the y-axis is in linear scale; for the lower panel, the y-axis is in logarithm scale.}}
	\label{f3}
\end{figure}

To illustrate the inherent diversity of initial growth rates, in Fig.~\ref{f3} (left panel), we display the daily cell-occupied areas plotted on a linear scale {and a logarithm scale} starting from Day 4. All wells with seed of $N_0=10$ or $N_0=4$ cells grew exponentially. Among the $N_0=1$-cell wells, 14 populations died out. Four wells in the $N_0=1$-cell group had more than 10 cells on Day 8 but never grew exponentially, and had fewer than 1000 cells after 15 days (on Day 23). For these non-growing or slow-growing $N_0=1$-cell wells, the per capita growth rate was $0\sim 0.2$. In comparison, all the $N_0=10$-cell wells needed at most 15 days to reach the carrying capacity (around 80 area units, or 40000 cells). See Table~\ref{gc} for a summary of the $N_0=1$-cell group's ``growth patterns''. This behavior is not idiosyncratic to the culture system because they recapitulate a pilot experiment performed in the larger scale format of 96-well plates (not shown).

\begin{table}[bt]
	\caption{\label{gc}The population of some wells in the $N_0=1$-cell group in the growth experiment with different initial cell numbers, where $\sim$ meant approximate cell number. These wells illustrated different growth patterns from those wells starting with $N_0=10$ or $N_0=4$ cells. Such differences implied that cells from wells with different initial cell numbers were essentially different.}

	\begin{tabular}{lllllll}
		\toprule
		Growth pattern                                                                                    & Well label                                                                                              & Day 1 & Day 8         & Day 14        & Day 19                                                       & Day 23\\    \midrule                                                   &&&&&&\\ 
		\begin{tabular}[c]{@{}l@{}}No growth, \\ extinction\end{tabular}                                  & \begin{tabular}[c]{@{}l@{}}162,167,170,176,\\ 177,179,182,183,\\ 186,201,234,236,\\ 239,240\end{tabular} & 1     & \textless{}10 & \textless{}10 & $\sim$0                                                      & Empty                                                        \\ &&&&&&\\ 
		\multirow{4}{*}{\begin{tabular}[c]{@{}l@{}}Slow growth, \\ no exponential \\ growth\end{tabular}} & 165                                                                                                      & 1     & 89            & $\sim$300     & $\sim$350                                                    & $\sim$500                                                    \\ 
		& 166                                                                                                      & 1     & 36            & $\sim$110     & $\sim$120                                                    & $\sim$150                                                    \\ 
		& 178                                                                                                      & 1     & 43            & $\sim$140     & $\sim$170                                                    & $\sim$200                                                    \\ 
		& 211                                                                                                      & 1     & 16            & $\sim$90      & $\sim$200                                                    & $\sim$400                                                    \\ &&&&&&\\ 
		\multirow{4}{*}{\begin{tabular}[c]{@{}l@{}}Delayed \\ exponential \\ growth\end{tabular}}         & 163                                                                                                      & 1     & 12            & $\sim$130     & $\sim$300                                                    & $\sim$5000                                                   \\  
		& 181                                                                                                      & 1     & 44            & $\sim$270     & $\sim$550                                                    & $\sim$5500                                                   \\ 
		& 193                                                                                                      & 1     & 25            & $\sim$200     & $\sim$800                                                    & $\sim$9000                                                   \\ 
		& 204                                                                                                      & 1     & 21            & $\sim$100     & $\sim$600                                                    & $\sim$6000                                                   \\ &&&&&&\\ 
		\begin{tabular}[c]{@{}l@{}}Normal \\ exponential \\ growth\end{tabular}                           & \begin{tabular}[c]{@{}l@{}}200 and\\ many others\end{tabular}                                            & 1     & $\sim$130     & $\sim$20000   & \begin{tabular}[c]{@{}l@{}}$\sim$40000\\ (full)\end{tabular} & \begin{tabular}[c]{@{}l@{}}$\sim$40000\\ (full)\end{tabular} \\ \bottomrule
	\end{tabular}

\end{table}

From the above experimental observations, we asserted for our model that there might be at least three stable cell growth phenotypes in a population: a fast type, whose growth rate was $0.6\sim 0.9/\text{day}$ for non-crowded conditions; a moderate type, whose growth rate was $0.3\sim 0.5/\text{day}$ for non-crowded conditions; and a slow type, whose growth rate was $0\sim 0.2/\text{day}$ for the non-crowded population. 

The graphs of Fig.~\ref{f3} also revealed other phenomena of growth kinetics: (1) Most $N_0=4$-cell wells plateaued by Day 14 to Day 17, but some lagged significantly behind. (2) Similarly, four wells in the $N_0=1$-cell group exhibited longer lag-times before the exponential growth phase, and never reached half-maximal cell numbers by Day 23. These outliers reveal intrinsic variability and were taken into account in the parameter scanning (see the Methods section).

\subsection{Reseeding experiments revealing the enduring intrinsic growth patterns.}

When a well in the $N_0=1$-cell group had grown to 10 cells, population behavior was still different from those in the $N_0=10$-cell group at the outset. In view of the spate of recent results revealing phenotypic heterogeneity, we hypothesized that the difference was cell-intrinsic as opposed to being a consequence of the environment (e.g., culture medium in the $N_0=1$ vs $N_0=10$ -cell wells, cell density). 

To test our hypothesis and exclude differences in the culture environment as determinants of growth behavior, we reseeded the cells that exhibited the different growth rates in fresh cultures. We started with a number of $N_0=1$-cell wells. After a period of almost 3 weeks, again some wells showed rapid proliferation, with cells covering the well, while others were half full and yet others wells were almost empty. {We divided the growth patterns into to groups by collecting cells either from the full (``faster wells'') or the half-full wells (``slower wells'') and} reseeded them into 32 wells each (at about $N_0=78$ cells per well). These 64 wells were monitored for another 20 days. We found that most wells reseeded from the ``faster wells'' took around 11 days to reach a given {endpoint population size (defined as filling the well by half)}, while most wells reseeded from the ``slower wells'' required around $16\sim 20$ days to reach the same endpoint population size. Five wells reseeded from the half-full wells were far from even reaching this endpoint population size by Day 20 (see Table~\ref{tnew}). Permutation test showed that this difference in growth rate was significant (see the Methods section).

This reseeding experiment shows that the difference in growth rate was maintained over multiple generations, even after slowing down in the plateau phase (full well) and was maintained when restarting a microculture at low density in fresh medium devoid of secreted cell products. Therefore, it is plausible that there exists endogenous heterogeneity of growth phenotypes in the clonal HL60 cell line and that these distinct growth phenotypes are stable for at least $15\sim 20$ cell generations.

\begin{table}[bt]
	\caption{\label{tnew}The distribution of time needed for each well to reach the ``half area'' population size in the reseeding experiment. We reseeded equal numbers of cells that grew faster (from a full well) and cells that grew slower (from a half-full well), and cultivated them under the same new fresh medium environment to compare their intrinsic growth rates. The results showed that faster growing cells, even reseeded, still grew faster.}
	\begin{tabular}{llllllll}
		\toprule
		\begin{tabular}[c]{@{}l@{}}Time (days) to \\reach one half area\end{tabular} & 11 & 12 & 13 & 14 & 15 & 16--20 & \textgreater20 \\ 
		\midrule
		Faster wells                                                            & 26 & 2  & 1  & 2  & 1  & 0  & 0              \\ 
		Slower wells                                                            & 0  & 0  & 0  & 1  & 1  & 25   &  5              \\ \bottomrule
	\end{tabular}

\end{table}

\subsection{Quantitative analysis of experimental results.}
In the experiments with different initial cell numbers $N_0$, we observed at least three patterns with different growth rates, and the reseeding showed that these growth patterns were endogenous to the cells. Therefore, we propose that each growth pattern discussed above corresponded to a cell phenotype that dominated the population, hereafter, referred to as: ``fast'', ``moderate'', and ``slow''.

In the initial seeding of cells for varying $N_0$, the cells were randomly chosen (by FACS); thus, their intrinsic growth phenotypes were randomly distributed. During growth, the population of a well would be dominated by the fastest type that existed in the seeding cells, thus qualitatively, we have following scenarios: (1) A well in the $N_0=10$-cell group almost certainly had at least one initial cell of fast type, and the population would be dominated by fast type cells. Different wells had almost the same growth rate, reaching saturation at almost the same time. (2) For an $N_0=1$-cell well, if the only initial cell is of the fast type, then the population has only the fast type, and the growth pattern will be close to that of $N_0=10$-cell wells. If the only initial cell is of the moderate type, then the population could still grow exponentially, but with a slower growth rate. This explains why after reaching 5 area units, many but not all $N_0=1$-cell wells were slower than $N_0=10$-cell wells. (3) Moreover, in such an $N_0=1$-cell well with a moderate type initial cell, the cell might not divide quite often during the first few days due to randomness of entering the cell cycle. This would lead to a considerable delay in entering the exponential growth phase. (4) By contrast, for an $N_0=1$-cell well with a slow type initial cell, the growth rate could be too small, and the population might die out or survive without ever entering the exponential growth phase in duration of the experiment. (5) Most $N_0=4$-cell wells had at least one fast type cell among the initial cells, and the growth pattern was the same as $N_0=10$-cell wells. A few $N_0=4$-cell wells only had moderate and slow cells, and thus had slower growth patterns. 

\begin{figure}[ht]
	\centering
	\includegraphics[width=0.9\linewidth]{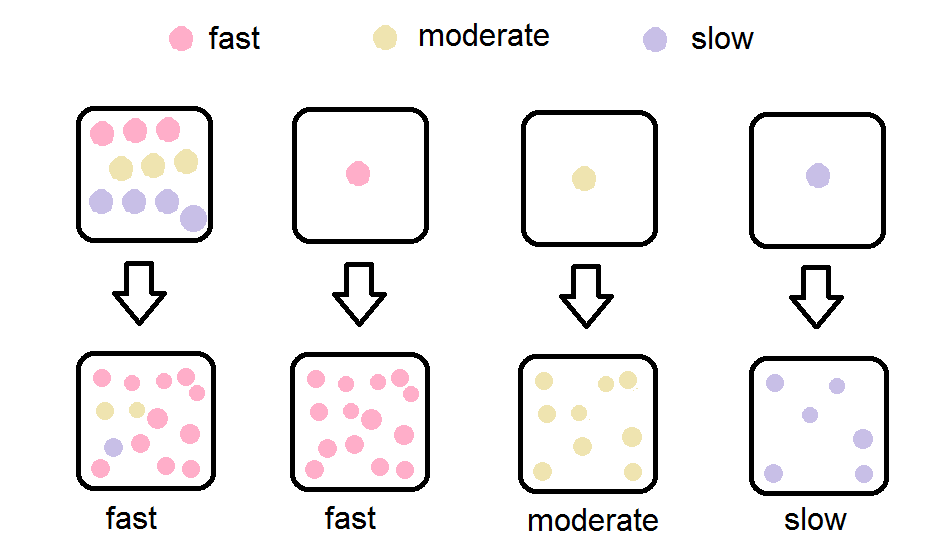}  
	\caption{Schematic illustration of the qualitative argument: Three cell types and growth patterns (three colors) with different seeding numbers. One $N_0=10$-cell well will have at least one fast type cell with high probability, which will dominate the population. One $N_0=1$-cell well can only have one cell type, thus in the microculture ensemble of replicate wells, three possible growth patterns for wells can be observed.}
	\label{f4}
\end{figure}

The above verbal argument is shown in Fig.~\ref{f4} and entails mathematical modeling with the appropriate parameters that relate the relative frequency of these cell types in the original population, their associated growth and transition rates to examine whether it explains the data. 

In the above analyses, we introduced fast and moderate types with different growth rates to explain the different growth rates for 10-cell wells and 1-cell wells. Another possible explanation for such a difference in growth rates was that the population would be 10 small colonies when starting from 10 initial cells, while starting from 1 initial cell, the population would be 1 large colony. With the same area, 10 small colonies should have a larger total perimeter, thus larger growth space and larger growth rate than that of 1 large colony. However, we carefully checked the photos, and found that almost all wells produced 1 large colony with nearly the same shape, and there was no significant relationship between colony perimeter and growth rate.

\subsection{Branching process model.}
To construct a quantitative dynamical model to recapitulate the growth dynamics differences from cell populations with distinct initial seed cell numbers $N_0$, and three intrinsic types of proliferation behaviors, we used a multi-type discrete-time branching process. 

The traditional method of population dynamics based on ordinary differential equation (ODE), which is deterministic and has continuous variables, is not suited when the cell population is small as is the case for the earliest stage of proliferation from a few cells being studied in our experiments. Deterministic models are also unfit because with such small populations and measurements at single-cell resolution, stochasticity in cell activity does not average out. The nuanced differences between individual cells cannot be captured by a different deterministic mechanism of each individual cell, and the only information available is the initial cell number. Thus, the unobservable nuances between cells are taken care of by a stochastic model.

Given the small populations, our model should be purely stochastic, without deterministic growth. The focus is the concrete population size of a finite number (three) of types, thus Poisson processes are not suitable. Markov chains can partially describe the proportions under some conditions, but population sizes are known, not just their ratios, therefore Markov chains are not necessary. Branching processes can describe the population size of multiple types with symmetric and asymmetric division, transitions between types, and death \cite{Yue}. Also, the parameters can be temporally and spatially inhomogeneous, which is convenient. Therefore, we utilized branching processes in our model. 

To simplify the simulation, we consider a discrete-time three-type branching process, which has been discussed in various literature \cite{harris1963theory,athreya2004branching,haccou2005branching,durrett2015branching,Yue}. After each time step, each living cell chooses a random behavior independently: to divide into two cells of the same type, to die, or to stay still. The probabilities of birth, death, and stagnation depend on the cell type and the current total cell number. Details of the model are described in the Methods Section below.







{We implemented the branching process model and ran simulations.} The simulation results are shown on the right panels of Figs.~\ref{f1}--\ref{f3}, in comparison with the experimental data in the left. Our model qualitatively captured the growth patterns of groups with different initial seeding cell numbers. For example, in Fig.~\ref{f2}, when wells were less than half full (cell number $<20000$), most wells in the $N_0=10$-cell group grew faster than the $N_0=1$-cell group even when they had the same cell number. In Fig.~\ref{f3}, all wells in the $N_0=10$-cell group in our model grew quickly until saturation. Similar to the experiment, some wells in the $N_0=1$-cell group in our model never grew, while some began to take off very late.

We used different methods to estimate the model parameters. In addition, we performed a parameter scan to show that our model could reproduce the experimentally observed phenomena for a wide range of model parameters (see details in Table~\ref{scan}).

\begin{table}[bt]
	\caption{\label{scan}Performance of our model with different parameters. Here we adjusted the parameters of our model in a wide range and observed whether the model could still reproduce four important ``features'' in the experiment. This parameter scan showed that our model is robust under perturbations on parameters. Here $p_\mathrm{F},p_\mathrm{M},p_\mathrm{S}$ are the probabilities that an initial cell is of fast, moderate, or slow type; $d$ is the death rate; $g_0$ is the growth factor; $r$ is the range of the random modifier. See the Methods section for explanations of these parameters. Feature 1, all wells in the $N_0=10$-cell group were saturated; Feature 2, presence of late-growing wells in the $N_0=1$-cell group; Feature 3, presence of non-growing wells in the $N_0=1$-cell group; Feature 4, different growth rates at the same population size between the $N_0=10$-cell group and the $N_0=1$-cell group.}
	\begin{tabular}{llllllllll}
		\toprule
		\multicolumn{6}{l}{Parameters}     & \multicolumn{4}{l}{Appearance of experimental phenomena} \\ 
		$p_\mathrm{F}$	&  $p_\mathrm{M}$   & $p_\mathrm{S}$    & $d$    & $g_0$    & $r$    & Feature 1    & Feature 2    & Feature 3    & Feature 4    \\ 
		\midrule
		0.4 & 0.4 & 0.2 & 0.01 & 0.5  & 0.1  & Yes          & Yes          & Yes          & Yes          \\ 
		0.4 & 0.4 & 0.2 & 0    & 0.5  & 0.1  & Yes          & Yes          & Yes          & Yes          \\ 
		0.4 & 0.4 & 0.2 & 0.05 & 0.5  & 0.1  & Yes          & Yes          & Yes          & Yes          \\ 
		0.4 & 0.4 & 0.2 & 0.1  & 0.5  & 0.1  & No           & Yes          & Yes          & No           \\ 
		0.4 & 0.4 & 0.2 & 0.01 & 0.45 & 0.1  & Yes          & Yes          & Yes          & Yes          \\ 
		0.4 & 0.4 & 0.2 & 0.01 & 0.6  & 0.1  & Yes          & Yes          & Yes          & Yes          \\ 
		0.4 & 0.4 & 0.2 & 0.01 & 0.4  & 0.1  & Yes          & Yes          & Yes          & No           \\ 
		0.4 & 0.4 & 0.2 & 0.01 & 0.5  & 0.05 & Yes          & Yes          & Yes          & Yes          \\ 
		0.4 & 0.4 & 0.2 & 0.01 & 0.5  & 0    & Yes          & Yes          & Yes          & Yes          \\ 
		0.4 & 0.4 & 0.2 & 0.01 & 0.5  & 0.15 & Yes          & Yes          & Yes          & No           \\ 
		0.4 & 0.4 & 0.2 & 0.01 & 0.5  & 0.2  & No           & Yes          & Yes          & No           \\ 
		0.3 & 0.5 & 0.2 & 0.01 & 0.5  & 0.1  & Yes          & Yes          & Yes          & Yes          \\ 
		0.5 & 0.3 & 0.2 & 0.01 & 0.5  & 0.1  & Yes          & Yes          & Yes          & Yes          \\ 
		0.4 & 0.5 & 0.1 & 0.01 & 0.5  & 0.1  & Yes          & Yes          & Yes          & Yes          \\ 
		0.4 & 0.3 & 0.3 & 0.01 & 0.5  & 0.1  & Yes          & Yes          & Yes          & Yes          \\ 
		0.5 & 0.4 & 0.1 & 0.01 & 0.5  & 0.1  & Yes          & Yes          & Yes          & Yes          \\ 
		0.3 & 0.4 & 0.3 & 0.01 & 0.5  & 0.1  & Yes          & Yes          & Yes          & Yes          \\ 
		0.1 & 0.1 & 0.8 & 0.01 & 0.5  & 0.1  & No           & Yes          & Yes          & No           \\ 
		\bottomrule
	\end{tabular}

\end{table}


In experiments, the $N_0=1$-cell group had a high extinction rate ($14/80=0.175 $). In the Methods section, we analyzed the extinction probability for our branching process model. For the $N_0=1$-cell group, the extinction probability is only $0.032$ in the model. We assume that the experimental procedure could cause damage for the seeding cells, as experience with single cell manipulation shows, so that the only initial cell for the $N_0=1$-cell group might be unhealthy or dead already at the beginning. Therefore, the extinction probability in experiments is higher.

\section{Discussion}
As many recent single-cell level data have shown, a tumor can contain multiple distinct subpopulations engaging in interconversions and interactions among them that can influence cancer cell proliferation, death, migration, and other features that contribute to malignancy \cite{pisco,zhou2014multi,angelini2022model,howard2018multi,sahoo2021mechanistic,zhou2014nonequilibrium,johnson2019cancer,korolev2014turning,chapman2014heterogeneous,niu2015phenotypic,chen2016overshoot}. Presence of these two intra-population behaviors can be manifest as departure from the elementary model of exponential growth \cite{skehan1984non} (in the early phase of population growth, far away from carrying capacity of the culture environment which is trivially non-exponential). The exponential growth model assumes uniformity of cell division rates across all cells (hence a population doubling rate that is proportional to a given population size $N(t)$) and the absence of cell-cell interactions that affect cell division and death rates. Investigating the ``non-genetic heterogeneity'' hypothesis of cancer cells quantitatively is therefore paramount for understanding cancer biology but also for elementary principles of cell population growth.

As an example, here we showed that clonal cell populations of the leukemia HL60 cell line are heterogeneous with regard to growth behaviors of individual cells that can be summarized in subpopulations characterized by a distinct intrinsic growth rates which were revealed by analysis of the early population growth starting with microcultures seeded with varying (low) cell number $N_0$. 

Since we have noted only very weak effect of cell-cell interactions on cell growth behaviors (Allee effect) in this cell line (as opposed to another cell tumor cell line in which we found that departure from exponential growth could be explained by the Allee effect \cite{johnson2019cancer}), we focused on the very presence among HL60 cells of subpopulations with distinct proliferative capacity as a mechanism for the departure of the early population growth curve from exponential growth. 

The reseeding experiment demonstrated that the characteristic growth behaviors of subpopulations could be inherited across cell generations and after moving to a new environment (fresh culture), consistent with long-enduring endogenous properties of the cells. This result might be explained by cells occupying distinct stable cell states (in a multi-stable system). Thus, we introduced multiple cell types with different growth rates in our stochastic model. Specifically, in a branching process model, we assumed the existence of three types: fast, moderate, and slow cells. The model we built could replicate the key features in the experimental data, such as different growth rates at the same population size between the $N_0=10$-cell group and the $N_0=1$-cell group, and the presence of late-growing and non-growing wells in the $N_0=1$-cell group.

While we were able to fit the observed behaviors in which the growth rate depended not only on $N(t)$ but also on $N_0$, the existence of the three or even more cell types still needs to be verified experimentally. For instance, statistical cluster analysis of transcriptomes of individual cells by single-cell RNA-seq \cite{bhartiya2021will} over the population may identify the presence of transcriptomically distinct subpopulations that could be isolated (e.g., after association with cell surface markers) and evaluated separately for their growth behaviors.

The central assumption of coexistence of multiple subpopulations in the cell line stock must be accompanied by the second assumption that there are transitions between these distinct cell populations. For otherwise, in the stock population the fastest growing cell would eventually outgrow the slow growing cells. Furthermore, one has to assume a steady-state in which the population of slow growing cells are continuously replenished from the population of fast-growing cells. Finally, we must assume that the steady-state proportions of the subpopulations are such that at low seeding wells with $N_0=1$ cells, there is a sizable probability that a microculture receives cells from each of the (three) presumed subtypes of cells. The number of wells in the ensemble of replicate microcultures for each $N_0$- condition has been sufficiently large for us to make the observations and inform the model, but a larger ensemble would be required to determine with satisfactory accuracy the relative proportions of the cell types in the parental stock population.

Transitions might also have been happening during our experiment. For example, those late growing wells in the $N_0=1$-cell group could be explained by such a transition: Initially, only slow type cells were present, but once one of these slow growing cells switched to the moderate type, an exponential growth ensued at the same rate that is intrinsic to that of moderate cells. 

If there are transitions, what is the transition rate? Our reseeding experiments are compatible with a relatively slow rate for interconversion of growth behaviors in that the same growth type was maintained across 30 generations. An alternative to the principle of transition at a constant intrinsic to each of the types of cells may be that transition is extrinsically determined. Specifically, the seeding in the ``lone'' condition of $N_0=1$ may \emph{induce} a dormant state, that is a transition to a slower growth mode that is then maintained, on average over 30+ generations, with occasional return to the faster types that account for the delayed exponential growth.

This model however would bring back the notion of ``environment awareness'', or the principle of a ``critical density'' for growth implemented by cell-cell interaction (Allee effect) which we had deliberately not considered (see above) since it was not necessary. We do not exclude this possibility which could be experimentally tested as follows: Cultivate $N_0=1$-cell wells for 20 days when the delayed exponential growth has happened in some wells, but then use the cells of those wells with fast-growing population (which should contain of the fast type) to restart the experiment, seeded at $N_0=10,4,1$ cells. If wells with different seeding numbers exhibit the same growth rates, then the growth difference in the original experiment is solely due to preexisting (slow interconverting) cell phenotypes. If now the $N_0=1$-cell wells resumes the typical slow growth, this would indicate a density induced transition to the slow growth type.

In the spirit of Occam's razor, and given the technical difficulty in separate experiments to demonstrate cell-cell interactions in HL60 cells, we were able to model the observed behaviors with the simplest assumption of cell-autonomous properties, including existence of multiple states (growth behaviors) and slow transitions between them but without cell density dependence or interactions.

Taken together, we showed that one manifestation of the burgeoning awareness of ubiquitous cell phenotype heterogeneity in an isogenic cell population is the presence of distinct intrinsic types of cells that slowly interconvert among them, resulting in a stationary population composition. The differing growth rates of the subtypes and their stable proportions may be an elementary characteristic of a given population that by itself can account for the departure of early population growth kinetics from the basic exponential growth model.

\section{Methods}
\subsection{Setup of growth experiment with different initial cell numbers.}
HL60 cells were maintained in IMDM wGln, 20\% FBS(heat inactivated), 1\% P/S at a cell density between $3\times 10^5$ and $2.5\times 10^6$ cells/ml (GIBCO). Cells were always handled and maintained under sterile conditions (tissue culture hood; $37^{\circ}$C, 5\% $\textrm{CO}_2$, humidified incubator). At the beginning of the experiment, cells were collected, washed two times in PBS, and stained for vitality (Trypan blue GIBCO). The population of cells was first gated for morphology and then for vitality staining. Only Trypan negative cells were sorted (BD FACSAria II). The cells were sorted in a 384 well plate with IMDM wGln, 20\% FBS(heat inactivated), and 1\% P/S (GIBCO). 

Cell population growth was monitored using a Leica microscope (heated environmental chamber and $\textrm{CO}_2$ levels control) with a motorized tray. Starting from Day 4, the 384 well plate was placed inside the environmental chamber every 24 hours. The images were acquired in a $3\times 3$ grid for each well; after acquisition, the 9 fields were stitched into a single image. Software ImageJ was applied to identify and estimate the area occupied by ``entities'' in each image. The area (proportional to cell number) was used to follow the cell growth.

\subsection{Setup of reseeding experiment for growth pattern inheritance.}
HL60 cells were cultivated for 3 weeks, and then we chose one full well and one half full well. We supposed the full well was dominated by fast type cells, and the half-full well was dominated by moderate type cells, which had lower growth rates. We reseeded cells from these two wells and cultivated them in two 96-well (rows A-H, columns 1-12) plates. In each plate, B2-B11, D2-D11, and F2-F11 wells started with 78 fast cells, while C2-C11, E2-E11, and G2-G11 wells started with 78 moderate cells. Rows A, H, columns 1, 12 had no cells and no media, and we found that wells in rows B, G, columns 2, 11, which were the outmost non-empty wells, evaporated much faster than inner wells. Therefore, the growth of cells in those wells was much slower than inner wells. Hence we only considered inner wells, where D3-D10 and F3-F10 started with fast cells, C3-C10 and E3-E10 started with moderate cells, namely 32 fast wells and 32 moderate wells in total. During the experiment, no media was added. Each day, we observed those wells to check whether their areas exceeded one-half of the whole well. The experiment was terminated after 20 days.

{
	\subsection{ANOVA and Weighted Welch's $t$-test.}
	
	The one-way ANOVA is used to test the hypothesis that multiple groups of samples all have equal mean \cite{st1989analysis}. Assume for group $i$ ($i=1,\ldots,k$), the sample size is $N_i$ and the $j$th sample is $X^j_i$. Set $N=\sum_{i=1}^k N_i$. Define 
	\[\bar{X}_i=\frac{1}{N_i}\sum_{j=1}^{N_i}X_i^j,\]
	\[\bar{X}=\frac{1}{N}\sum_{i=1}^k\sum_{j=1}^{N_i}X_i^j,\]
	\[F=\frac{\sum_{i=1}^k N_i(\bar{X}_i-\bar{X})^2/(k-1)}{\sum_{i=1}^k \sum_{j=1}^{N_i} (X_i^j-\bar{X}_i)^2/(N-k)}.\]
	If groups $i=1,\ldots,k$ all have the same mean, then $F$ should follow the $F$-distribution.
	
	The one-way ANOVA was applied to the growth experiment with different initial cell numbers, in order to determine whether the growth rates during the exponential phase (5--50 area units) were different among all three groups. Here $X^j_i$ corresponded to the growth rate. The $p$-value was $1.77\times 10^{-25}$. Therefore, the growth rate difference among all three groups was statistically significant.
	
	In ANOVA, we treated all $X_i^j$ with the same weight. However, in the experiments, the growth rate $X_i^j$ is the average of growth rates for all cells. Thus the cell number (area) should be the weight. To account for different weights, we also applied the weighted Welch's $t$-test, although it only compares two populations.
}

The weighted Welch's $t$-test is used to test the hypothesis that two populations have equal mean, while sample values have different weights \cite{weight}. Assume for group $i$ ($i=1,2$), the sample size is $N_i$ and the $j$th sample is the average of $c^j_i$ independent and identically distributed variables. Let $X^j_i$ be the observed average for the $j$th sample. Set $\nu_1=N_1-1$, $\nu_2=N_2-1$. Define 
\begin{equation*}
	\begin{split}
		&\bar{X_i}^W=(\sum_{j=1}^{N_i}X^j_i c_j)/(\sum_{j=1}^{N_i})c_j,\\
		&s_{i,W}^2=\frac{N_i[\sum_{j=1}^{N_i}(X^j_i)^2 c_j]/(\sum_{j=1}^{N_i}c_i^j)-N_i(\bar{X_i}^W)^2}{N_i-1},\\
		&t=\frac{\bar{X_1}^W-\bar{X_2}^W}{\sqrt{\frac{s_{1,W}^2}{N_1}+\frac{s_{2,W}^2}{N_2}}},\\
		&\nu=\frac{(\frac{s_{1,W}^2}{N_1}+\frac{s_{2,W}^2}{N_2})^2}{\frac{s_{1,W}^4}{N_1^2\nu_1}+\frac{s_{2,W}^4}{N_2^2\nu_2}}.
	\end{split}
\end{equation*}
If two populations have equal mean, then $t$ satisfies the $t$-distribution with degree of freedom $\nu$.

The weighted Welch's $t$-test was applied to the growth experiment with different initial cell numbers, in order to determine whether the growth rates during exponential phase (5--50 area units) were different between groups. Here $X^j_i$ corresponded to growth rate, and $c^j_i$ corresponded to cell area. The $p$-value for $N_0=10$-cell group vs. $N_0=4$-cell group was $2.12\times 10^{-8}$; the $p$-value for $N_0=10$-cell group vs. $N_0=1$-cell group was smaller than $10^{-12}$; the $p$-value for $N_0=4$-cell group vs. $N_0=1$-cell group was $5.35\times 10^{-5}$. Here we conducted three tests. To avoid increasing the probability of type I error \cite{sokal1981biometry}, we performed the Bonferroni correction \cite{armstrong2014use}. This means that the significance level $\alpha$ should be divided by the number of tests. In our situation, there are $3$ tests, and each $p$-value is smaller than $2\times 10^{-4}/3$, meaning that the difference is significant for the level $\alpha=2\times 10^{-4}$. Therefore, the growth rate difference between any two groups was statistically significant.

\subsection{Permutation Test.}
The permutation test is a non-parametric method to test whether two samples are significantly different with respect to a statistic (e.g., sample mean) \cite{perm}. It is easy to calculate and fits our situation, thus we adopt this test rather than other more complicated tests, such as the Mann-Whitney test. For two samples $\{x_1,\cdots,x_m\}$, $\{y_1,\cdots,y_n\}$, consider the null hypothesis: the mean of $x$ and $y$ are the same. For these samples, calculate the mean of the first sample: $\mu_0=\frac{1}{m}\sum x_i$. Then we randomly divide these $m+n$ samples into two groups with size $m$ and $n$: $\{x_1',\cdots,x_m'\}$, $\{y_1',\cdots,y_n'\}$, such that each permutation has equal probability. For these new samples, calculate the mean of the first sample: $\mu_0'=\frac{1}{m}\sum x_i'$. Then the two-sided $p$-value is defined as 
\[p=2\min \{ \mathbb{P}(\mu_0\le \mu_0'),1-\mathbb{P}(\mu_0\le \mu_0')\}.\]
If $\mu_0$ is an extreme value in the distribution of $\mu_0'$, then the two sample means are different. 

In the reseeding experiment, the mean time of exceeding half well for the fast group was 11.4375 days. For all $\binom{64}{32}$ possible result combinations, only 7 combinations had equal or less mean time. Thus the $p$-value was $2\times7/\binom{64}{32}=7.6\times 10^{-18}$. This indicated that the growth rate difference between fast group and moderate group was significant.

\subsection{Model Details.}
We built a branching process model to describe the population growth dynamics of populations containing a mixture of fast, moderate, and slow type cells. In our model, the branching process is discrete in time, and each time step is half a day. The state space of the branching process is $[F(t),M(t),S(t)]\in (\mathbb{Z}^*)^3$, representing the number of fast-type cells, the number of moderate-type cells, and the number of slow-type cells at time $t$. Here $t$ can take $0,0.5,1,1.5,\ldots$. This is a standard multi-type discrete-time branching process \cite{harris1963theory,athreya2004branching,haccou2005branching,durrett2015branching,Yue}, although the transition probabilities can depend on the current state, thus being time-inhomogeneous.

At time $t=0$, there are $N_0$ initial cells ($N_0=10/4/1$ in the experiments). For each initial cell, the type is independently chosen, and the probabilities of being fast, moderate, or slow type are $p_\mathrm{F},p_\mathrm{M},p_\mathrm{S}$, where $p_\mathrm{F}+p_\mathrm{M}+p_\mathrm{S}=1$. Therefore, the type distribution of the initial population is
\[\mathbb{P}[F(0)=a,M(0)=b,S(0)=c]=\frac{N_0!}{a!b!c!}a^{p_\mathrm{F}}b^{p_\mathrm{M}}c^{p_\mathrm{S}},\]
where $a+b+c=N_0$.

The branching process evolves as follows: After each time step (half a day), each cell independently and randomly chooses a behavior: division, death, or stagnation in the quiescent state, whose probabilities depend on the cell type and the current total population. Therefore, given the population at time $t$, $[F(t),M(t),S(t)]$, the population at $t+0.5$ (the next time point) is
\[[F(t+0.5),M(t+0.5),S(t+0.5)]=[\sum_{i=1}^{F(t)}A_i,\sum_{j=1}^{M(t)}B_j,\sum_{k=1}^{S(t)}C_k].\]
Here $A_i,B_j,C_k$ are independent random variables. Different $A_i$ have the same distribution:
\[\mathbb{P}(A_i=0)=d_\mathrm{F}(t),\ \mathbb{P}(A_i=1)=1-g_\mathrm{F}(t)-d_\mathrm{F}(t), \ \mathbb{P}(A_i=2)=g_\mathrm{F}(t),\]
where $g_\mathrm{F}(t)$ and $d_\mathrm{F}(t)$ are the birth and death probabilities for fast type cells at time $t$, defined later. Similarly, for moderate type cells, different $B_j$ have the same distribution:
\[\mathbb{P}(B_j=0)=d_\mathrm{M}(t),\ \mathbb{P}(B_j=1)=1-g_\mathrm{M}(t)-d_\mathrm{M}(t), \ \mathbb{P}(B_j=2)=g_\mathrm{M}(t),\]
with birth and death probabilities $g_\mathrm{M}(t), d_\mathrm{M}(t)$. For slow type cells, different $C_k$ have the same distribution:
\[\mathbb{P}(C_k=0)=d_\mathrm{S}(t),\ \mathbb{P}(C_k=1)=1-g_\mathrm{S}(t)-d_\mathrm{S}(t), \ \mathbb{P}(C_k=2)=g_\mathrm{S}(t),\]
with birth and death probabilities $g_\mathrm{S}(t), d_\mathrm{S}(t)$.

Therefore, given $[F(t),M(t),S(t)]$, the distribution of $[F(t+0.5),M(t+0.5),S(t+0.5)]$ is:
\begin{equation*}
	\begin{split}
		&\mathbb{P}[F(t+0.5)=N_1,M(t+0.5)=N_2,S(t+0.5)=N_3]\\
		=&\Big\{\sum_{2a+b=N_1}\frac{F(t)!}{a!b![F(t)-a-b]!}g_\mathrm{F}^a d_\mathrm{F}^{[F(n)-a-b]}(1-g_\mathrm{F}-d_\mathrm{F})^b\Big\}\\
		\times&\Big\{\sum_{2c+d=N_2}\frac{M(t)!}{c!d![M(t)-c-d]!}g_\mathrm{M}^c d_\mathrm{M}^{[M(t)-c-d]}(1-g_\mathrm{M}-d_\mathrm{M})^d\Big\}\\
		\times&\Big\{\sum_{2e+f=N_3}\frac{S(t)!}{e!f![S(t)-e-f]!}g_\mathrm{S}^e d_\mathrm{S}^{[S(t)-e-f]}(1-g_\mathrm{S}-d_\mathrm{S})^f\Big\},
	\end{split}
\end{equation*}
where the first summation is taken for all non-negative integer pairs $(a,b)$ with $2a+b=N_1$, and the other two summations are defined similarly.

In the model, the death probabilities are the same constant, $d_\mathrm{F}(t)=d_\mathrm{M}(t)=d_\mathrm{S}(t)=d$ for any $t$.

For the birth probabilities $g_\mathrm{F}(t),g_\mathrm{M}(t),g_\mathrm{S}(t)$, {\color{black}to determine their functional forms as function of the population size, we fitted the data shown in Fig.~\ref{f2}. Note that the $y$-axis of Fig.~\ref{f2} is the per capita growth rate, which is approximately two times the birth probability. Therefore, we took the average of the data points in Fig.~\ref{f2} with respect to the $y$-axis to obtain an approximated curve of the birth probabilities as a function of the population size, and fit this averaged curve (not shown) with different forms of functions to determine the best form for the birth probabilities. We tested three forms of functions, all with two parameters: $g=a+bN$ (linear), $g=a+bN^2$ (quadratic), and $g=a+b \log(N)$ (logarithmic). For each function form, we performed regression analysis to determine the parameter values, and calculated the mean square error. Since the quadratic form has the smallest error, we set }
\[g_\mathrm{F}(t)=g_0\{1-[F(t)+M(t)+S(t)]^2/C^2\}+\delta.\] 
Here $g_0$ is constant, representing the base growth probability; $C$ is a constant, representing the carrying capacity; $\delta$ is a random variable that satisfies the uniform distribution on $[-r,r]$ with a constant $r$, and $\delta$ is the same for any $t$ in the same simulation, namely all cells in the same well. If $g_\mathrm{F}(t)<0$, set $g_\mathrm{F}(t)=0$. If $g_\mathrm{F}(t)>1-d$, set $g_\mathrm{F}(t)=1-d$. After determining $g_\mathrm{F}(t)$, set $g_\mathrm{M}(t)=g_\mathrm{F}(t)/1.5$, and $g_\mathrm{S}(t)=g_\mathrm{F}(t)/3$.

There are seven parameters that need to be determined. In the simulations displayed, initial type probabilities $p_\mathrm{F}=0.4$, $p_\mathrm{M}=0.4$, $p_\mathrm{S}=0.2$, death rate $d=0.01$, carrying capacity $C=40000$, growth factor $g_0=0.5$, and the range of random modifier $r=0.1$. See Subsection~\ref{estimation} for how these parameters were chosen.

\subsection{Extinction Probability.}
\label{ext}
One important term in branching processes is the extinction probability $\gamma$, namely the probability that the population becomes zero within finite time. In our model, we can calculate the extinction probability explicitly from the model parameters.

We first consider a simple branching process model. There is only one cell type, where one cell has probability $g$ to divide and probability $d$ to die in one time step. Here $g$ and $d$ are constants. At the same time, different cells are independent. We use $X(t)=N$ to denote that the population size is $N$ at time step $t$. Assume the initial population is $X(0)=N_0$. In this model, the extinction probability is well known \cite{athreya2004branching}:

(1) If $g\le d$, 
\[\gamma=\lim_{t\to \infty}\mathbb{P}[X(t)=0]=1.\]

(2) If $g>d$, 
\[\gamma=\lim_{t\to \infty}\mathbb{P}[X(t)=0]=\left(\frac{d}{g}\right)^{N_0}.\]

In our model, for each cell type, the division probability slowly decreases with the total population. Since extinction is very unlikely when the total population is large, we can use the following approximation to study extinction:

\[g_\mathrm{F} \approx g_0+\delta.\]

Under this approximation, again, different cells are independent. If the process starts with a fast type cell, the extinction probability is $\gamma_\mathrm{F}=d/g_\mathrm{F}$. Since $\delta$ is uniform on $[-r,r]$, the expected extinction probability is
\[\mathbb{E}(\gamma_\mathrm{F})=\mathbb{E}[d/(g_0+\delta)]=\frac{d}{2r}\log\frac{g_0+r}{g_0-r}\approx \frac{d}{g_0}.\]
The last step uses the Taylor expansion $\log(1+x)\approx x$.

In our simulations, we set $d=0.01$, $g_0=0.5$. Thus $\gamma_\mathrm{F}=0.02$.

Similarly, if the process starts with one moderate type cell or a slow type cell, the expected extinction probability is 
\[\mathbb{E}(\gamma_\mathrm{M})\approx \frac{3d}{2g_0}=0.03.\]
\[\mathbb{E}(\gamma_\mathrm{S})\approx \frac{3d}{g_0}=0.06.\]

Notice that the initial cell has probability $p_\mathrm{F}=0.4$ to be fast type, $p_\mathrm{M}=0.4$ to be fast type, and $p_\mathrm{S}=0.2$ to be fast type. Therefore, if the initial population size is $1$, the overall extinction probability $\gamma$ is 
\[\gamma= p_\mathrm{F}\gamma_\mathrm{F}+p_\mathrm{M}\gamma_\mathrm{M}+p_\mathrm{S}\gamma_\mathrm{S},\]
and its expectation is
\[\mathbb{E}(\gamma)= p_\mathrm{F}\mathbb{E}(\gamma_\mathrm{F})+p_\mathrm{M}\mathbb{E}(\gamma_\mathrm{M})+p_\mathrm{S}\mathbb{E}(\gamma_\mathrm{S})=0.032.\]

If the process starts with $N_0>1$ cells, the extinction probability is $\gamma^{N_0}$. We can use the approximation $\mathbb{E}(\gamma^{N_0})\approx [\mathbb{E}(\gamma)]^{N_0}$. Therefore, if the process starts with $4$ cells or $10$ cells, the expected extinction probability is $0.032^4\approx 1.05\times 10^{-6}$ or $0.032^{10}\approx 1.13\times 10^{-15}$, which are both negligible.

\subsection{Parameter estimation.}	
\label{estimation}
In our model, seven main parameters need to be determined: initial type probabilities $p_\mathrm{F},p_\mathrm{M},p_\mathrm{S}$, death rate $d$, carrying capacity $C$, growth factor $g_0$, and the range of random modifier $r$.

We had separately determined that one area unit equals approximately 500 cells. When one well was full, the area was about 80 units. Therefore, we chose $C=40000$.

To determine initial type probabilities $p_\mathrm{F},p_\mathrm{M},p_\mathrm{S}$, we considered $N_0=1$-cell wells, since each of them had only one cell type. In those $80$ wells, $18$ wells never reached exponential growth phase, meaning that they might start with a slow type cell. When the cell area was small, we observed that fast type cells had growth rates $0.6\sim 0.9$, and moderate type cells had growth rates $0.3\sim 0.5$. Therefore, we calculated the mean growth rate for each well when the cell area was between $5$ and $20$, and compared the growth rate with a threshold $0.55$. Here area data under $5$ were not very accurate, and the growth rate started to decrease when the area was over $20$. For $62$ wells that ever reached exponential growth phase, $36$ wells had growth rates larger than $0.55$, and $26$ wells had growth rates smaller than $0.55$. If we set the threshold to be $0.6$, then $28$ wells had growth rates larger than $0.6$, and $34$ wells had growth rates smaller than $0.6$. Overall, we set $p_\mathrm{S}=0.2$, which is close to $18/80$. We observed that $p_\mathrm{F}$ and $p_\mathrm{M}$ should be close. Thus we set them as $p_\mathrm{F}=0.4,p_\mathrm{M}=0.4$.

To determine the death rate $d$, growth factor $g_0$, and the range of random modifier $r$, we calculated the maximum likelihood estimation. Since calculating the probability of generating certain data with given parameters requires accurate cell numbers we chose $10$ fast $1$-cell wells and counted the cell number manually for days when the cell number was smaller than $300$. For cells in one well, the actual growth rate when the cell number is not too large was $g_0+\delta$, where $\delta$ is uniform on $[-r,r]$, and the death rate was $d$. These rates were for half a day. Therefore, for one cell, the probabilities for generating $0,1,2,3,4$ cells after one day were:
\begin{equation*}
	\begin{split}
		&\mathbb{P}(N=1\to N=0\mid g,d)=gd^2+d(1-g-d)+d,\\
		&\mathbb{P}(N=1\to N=1\mid g,d)=2gd(1-g-d)+(1-g-d)^2,\\
		&\mathbb{P}(N=1\to N=2\mid g,d)=g(1-g-d)^2+2g^2d+g(1-g-d),\\
		&\mathbb{P}(N=1\to N=3\mid g,d)=2g^2(1-g-d),\\
		&\mathbb{P}(N=1\to N=4\mid g,d)=g^3.
	\end{split}
\end{equation*}
Given the above probabilities, we could calculate the probability that the cell number grew from $N=n_1$ cells to $N=n_2$ cells in one day, $\mathbb{P}(N=n_1\to N=n_2\mid g,d)$. Therefore, we had the probability of generating the cell numbers $N^i=(n_1^i,n_2^i,\ldots,n_k^i)$ in $k$ consecutive days for well $i$: 
\[\mathbb{P}(N^i\mid g,d)=\prod_{j=1}^{k-1}\mathbb{P}(N=n_j^i\to N=n_{j+1}^i\mid g,d).\]
Given $g_0,r,d$, we had the conditional probability 
\[\mathbb{P}(g=g^*,d\mid g_0,r,d)=\mathbb{P}(g_0,r,d,g=g^*\mid g_0,r,d)=\mathbb{P}(g=g^*\mid g_0,r).\] 
Therefore, the conditional probability of generating data $N^i$ with parameters $g_0,r,d$ is
\[\mathbb{P}(N^i\mid g_0,d,r)=\sum_{g^*}\mathbb{P}(N^i\mid g=g^*,d)\mathbb{P}(g=g^*,d\mid g_0,r,d).\]
Last, we had the conditional probability of generating data $N^1,\cdots,N^{10}$ with parameters $g_0,r,d$:
\begin{equation*}
	\begin{split}
		&\mathbb{P}(N^1,\cdots,N^{10}\mid g_0,d,r)\\
		=&\prod_{i=1}^{10}\mathbb{P}(N^i\mid g_0,d,r)\\
		=&\prod_{i=1}^{10}\left[\sum_{g^*}\mathbb{P}(N^i\mid g=g^*,d)\mathbb{P}(g=g^*,d\mid g_0,r,d)\right].
	\end{split}
\end{equation*}

We tested different combinations of $g_0,r,d$, and found the one that corresponded to the largest $\mathbb{P}(N^1,\cdots,N^{10}\mid g_0,d,r)$: $g_0=0.50$, $d=0.00$, $r=0.09$. Since we did not want a zero death rate, we set $g_0=0.5$, $d=0.01$, and $r=0.1$ in simulations.

\subsection{Parameter scan.}
Since we did not experimentally determine the cell type, and we did not have accurate cell number for most wells, the above parameter estimations may not be accurate either. Therefore, we performed a parameter scan by evaluating the performance of our model for different sets of parameters. {Except for the carrying capacity $C$ that was determined accurately, we adjusted six main parameters}: initial type probabilities $p_\mathrm{F}$, $p_\mathrm{M}$, $p_\mathrm{S}$, death rate $d$, growth factor $g_0$, and random modifier $r$. We checked whether these 4 features observable in the experiment could be reproduced: growth of all wells in the $N_0=10$-cell group to saturation; existence of late-growing wells in the $N_0=1$-cell group; existence of non-growing wells in the $N_0=1$-cell group; difference in growth rates in the $N_0=10$-cell group and the $N_0=1$-cell group at the same population size. Table~\ref{scan} shows the results of the performance of simulations with the various parameter sets. Within a wide range of parameters, our model is able to replicate the experimental results shown in Figs.~\ref{f1}--\ref{f3}, indicating that our model is robust under perturbations.

\section*{Acknowledgements}
This research was partially supported by NIGMS NIH-R01CA226258-01. We would like to thank Ivana Bozic, Yifei Liu, Georg Luebeck, Weili Wang, Yuting Wei and Lingxue Zhu for helpful advice and discussions.

\section*{Data availability}
The experimental data, simulation data, and corresponding code files could be found at https://github.com/YueWangMathbio/Leukemia.

\bibliographystyle{acm}
\bibliography{Single_Cell_Growth}

\begin{thebibliography}{10}

\bibitem{angelini2022model}
{\sc Angelini, E., Wang, Y., Zhou, J.~X., Qian, H., and Huang, S.}
\newblock A model for the intrinsic limit of cancer therapy: {Duality} of
  treatment-induced cell death and treatment-induced stemness.
\newblock {\em PLOS Comput. Biol. 18}, 7 (2022), e1010319.

\bibitem{armstrong2014use}
{\sc Armstrong, R.~A.}
\newblock When to use the bonferroni correction.
\newblock {\em Ophthalmic and Physiological Optics 34}, 5 (2014), 502--508.

\bibitem{athreya2004branching}
{\sc Athreya, K.~B., and Ney, P.~E.}
\newblock {\em Branching processes}.
\newblock Courier Corporation, 2004.

\bibitem{AS81}
{\sc Bartoszynski, R., Brown, B.~W., McBride, C.~M., and Thompson, J.~R.}
\newblock Some nonparametric techniques for estimating the intensity function
  of a cancer related nonstationary {Poisson} process.
\newblock {\em Ann. Stat.\/} (1981), 1050--1060.

\bibitem{bhartiya2021will}
{\sc Bhartiya, D., Kausik, A., Singh, P., and Sharma, D.}
\newblock Will single-cell rnaseq decipher stem cells biology in normal and
  cancerous tissues?
\newblock {\em Hum. Reprod. Update 27}, 2 (2021), 421--421.

\bibitem{chang2008transcriptome}
{\sc Chang, H.~H., Hemberg, M., Barahona, M., Ingber, D.~E., and Huang, S.}
\newblock Transcriptome-wide noise controls lineage choice in mammalian
  progenitor cells.
\newblock {\em Nature 453}, 7194 (2008), 544--547.

\bibitem{chapman2014heterogeneous}
{\sc Chapman, A., del Ama, L.~F., Ferguson, J., Kamarashev, J., Wellbrock, C.,
  and Hurlstone, A.}
\newblock Heterogeneous tumor subpopulations cooperate to drive invasion.
\newblock {\em Cell Rep. 8}, 3 (2014), 688--695.

\bibitem{chen2016overshoot}
{\sc Chen, X., Wang, Y., Feng, T., Yi, M., Zhang, X., and Zhou, D.}
\newblock The overshoot and phenotypic equilibrium in characterizing cancer
  dynamics of reversible phenotypic plasticity.
\newblock {\em J. Theor. Biol. 390\/} (2016), 40--49.

\bibitem{Clark1991}
{\sc Clark, W.~H.}
\newblock Tumour progression and the nature of cancer.
\newblock {\em Br. J. Cancer 64}, 4 (1991), 631.

\bibitem{dewanji}
{\sc Dewanji, A., Luebeck, E.~G., and Moolgavkar, S.~H.}
\newblock A generalized {Luria}--{Delbr{\"u}ck} model.
\newblock {\em Math. Biosci. 197}, 2 (2005), 140--152.

\bibitem{durrett2015branching}
{\sc Durrett, R.}
\newblock {\em Branching process models of cancer}.
\newblock Springer, 2015.

\bibitem{durrett2011intratumor}
{\sc Durrett, R., Foo, J., Leder, K., Mayberry, J., and Michor, F.}
\newblock Intratumor heterogeneity in evolutionary models of tumor progression.
\newblock {\em Genetics 188}, 2 (2011), 461--477.

\bibitem{Egeblad2010}
{\sc Egeblad, M., Nakasone, E.~S., and Werb, Z.}
\newblock Tumors as organs: {Complex} tissues that interface with the entire
  organism.
\newblock {\em Dev. Cell 18}, 6 (2010), 884--901.

\bibitem{Gatenby2014}
{\sc Gatenby, R.~A., Cunningham, J.~J., and Brown, J.~S.}
\newblock Evolutionary triage governs fitness in driver and passenger mutations
  and suggests targeting never mutations.
\newblock {\em Nat. Commun. 5\/} (2014), 5499.

\bibitem{weight}
{\sc Goldberg, L.~R., Kercheval, A.~N., and Lee, K.}
\newblock T-statistics for weighted means in credit risk modeling.
\newblock {\em J. Risk Finance 6}, 4 (2005), 349--365.

\bibitem{gunnarsson2020understanding}
{\sc Gunnarsson, E.~B., De, S., Leder, K., and Foo, J.}
\newblock Understanding the role of phenotypic switching in cancer drug
  resistance.
\newblock {\em J. Theor. Biol 490\/} (2020), 110162.

\bibitem{Gupta}
{\sc Gupta, P.~B., Fillmore, C.~M., Jiang, G., Shapira, S.~D., Tao, K.,
  Kuperwasser, C., and Lander, E.~S.}
\newblock Stochastic state transitions give rise to phenotypic equilibrium in
  populations of cancer cells.
\newblock {\em Cell 146}, 4 (2011), 633--644.

\bibitem{haccou2005branching}
{\sc Haccou, P., Jagers, P., and Vatutin, V.~A.}
\newblock {\em Branching processes: variation, growth, and extinction of
  populations}.
\newblock No.~5. Cambridge university press, 2005.

\bibitem{hanahan}
{\sc Hanahan, D., and Weinberg, R.~A.}
\newblock Hallmarks of cancer: {The} next generation.
\newblock {\em Cell 144}, 5 (2011), 646--674.

\bibitem{harris1963theory}
{\sc Harris, T.~E., et~al.}
\newblock {\em The theory of branching processes}, vol.~6.
\newblock Springer Berlin, 1963.

\bibitem{perm}
{\sc Hastie, T., Tibshirani, R., and Friedman, J.}
\newblock {\em The Elements of Statistical Learning}, 2nd~ed.
\newblock Springer, New York, 2016.

\bibitem{hordijk2018autocatalytic}
{\sc Hordijk, W., Steel, M., and Dittrich, P.}
\newblock Autocatalytic sets and chemical organizations: modeling
  self-sustaining reaction networks at the origin of life.
\newblock {\em New J. Phys. 20}, 1 (2018), 015011.

\bibitem{howard2018multi}
{\sc Howard, G.~R., Johnson, K.~E., Rodriguez~Ayala, A., Yankeelov, T.~E., and
  Brock, A.}
\newblock A multi-state model of chemoresistance to characterize phenotypic
  dynamics in breast cancer.
\newblock {\em Sci. Rep. 8}, 1 (2018), 1--11.

\bibitem{Yue}
{\sc Jiang, D.-Q., Wang, Y., and Zhou, D.}
\newblock Phenotypic equilibrium as probabilistic convergence in
  multi-phenotype cell population dynamics.
\newblock {\em PLOS ONE 12}, 2 (2017), e0170916.

\bibitem{johnson2019cancer}
{\sc Johnson, K.~E., Howard, G., Mo, W., Strasser, M.~K., Lima, E.~A., Huang,
  S., and Brock, A.}
\newblock Cancer cell population growth kinetics at low densities deviate from
  the exponential growth model and suggest an {Allee} effect.
\newblock {\em PLOS Biol. 17}, 8 (2019), e3000399.

\bibitem{Koch}
{\sc Koch, A.~L.}
\newblock Mutation and growth rates from {L}uria-{D}elbr{\"u}ck fluctuation
  tests.
\newblock {\em Mutat. Res. -Fund. Mol. M. 95}, 2-3 (1982), 129--143.

\bibitem{kochanowski2021systematic}
{\sc Kochanowski, K., Sander, T., Link, H., Chang, J., Altschuler, S.~J., and
  Wu, L.~F.}
\newblock Systematic alteration of in vitro metabolic environments reveals
  empirical growth relationships in cancer cell phenotypes.
\newblock {\em Cell Rep. 34}, 3 (2021), 108647.

\bibitem{korolev2014turning}
{\sc Korolev, K.~S., Xavier, J.~B., and Gore, J.}
\newblock Turning ecology and evolution against cancer.
\newblock {\em Nat. Rev. Cancer 14}, 5 (2014), 371--380.

\bibitem{Lea}
{\sc Lea, D.~E., and Coulson, C.~A.}
\newblock The distribution of the numbers of mutants in bacterial populations.
\newblock {\em J. Genet. 49}, 3 (1949), 264--285.

\bibitem{li2016dynamics}
{\sc Li, Q., Wennborg, A., Aurell, E., Dekel, E., Zou, J.-Z., Xu, Y., Huang,
  S., and Ernberg, I.}
\newblock Dynamics inside the cancer cell attractor reveal cell heterogeneity,
  limits of stability, and escape.
\newblock {\em Proc. Natl. Acad. Sci. U.S.A. 113}, 10 (2016), 2672--2677.

\bibitem{Luebeck}
{\sc Luebeck, G., and Moolgavkar, S.~H.}
\newblock Multistage carcinogenesis and the incidence of colorectal cancer.
\newblock {\em Proc. Natl. Acad. Sci. 99}, 23 (2002), 15095--15100.

\bibitem{Luria}
{\sc Luria, S.~E., and Delbr{\"u}ck, M.}
\newblock Mutations of bacteria from virus sensitivity to virus resistance.
\newblock {\em Genetics 28}, 6 (1943), 491.

\bibitem{Mackillop1990}
{\sc Mackillop, W.~J.}
\newblock The growth kinetics of human tumours.
\newblock {\em Clin. Phys. Physiol. M. 11}, 4A (1990), 121.

\bibitem{meacham_tumour_2013}
{\sc Meacham, C.~E., and Morrison, S.~J.}
\newblock Tumour heterogeneity and cancer cell plasticity.
\newblock {\em Nature 501}, 7467 (2013), 328--337.

\bibitem{lung}
{\sc Newton, P.~K., Mason, J., Bethel, K., Bazhenova, L.~A., Nieva, J., and
  Kuhn, P.}
\newblock A stochastic markov chain model to describe lung cancer growth and
  metastasis.
\newblock {\em PLOS ONE 7}, 4 (2012), e34637.

\bibitem{niu2015phenotypic}
{\sc Niu, Y., Wang, Y., and Zhou, D.}
\newblock The phenotypic equilibrium of cancer cells: {From} average-level
  stability to path-wise convergence.
\newblock {\em J. Theor. Biol. 386\/} (2015), 7--17.

\bibitem{pisco}
{\sc Pisco, A.~O., and Huang, S.}
\newblock Non-genetic cancer cell plasticity and therapy-induced stemness in
  tumour relapse: `{What} does not kill me strengthens me'.
\newblock {\em Br. J. Cancer 112}, 11 (2015), 1725--1732.

\bibitem{sahoo2021mechanistic}
{\sc Sahoo, S., Mishra, A., Kaur, H., Hari, K., Muralidharan, S., Mandal, S.,
  and Jolly, M.~K.}
\newblock A mechanistic model captures the emergence and implications of
  non-genetic heterogeneity and reversible drug resistance in {ER+} breast
  cancer cells.
\newblock {\em NAR Cancer 3}, 3 (2021), zcab027.

\bibitem{skehan1984non}
{\sc Skehan, P., and Friedman, S.~J.}
\newblock Non-exponential growth by mammalian cells in culture.
\newblock {\em Cell Prolif. 17}, 4 (1984), 335--343.

\bibitem{sokal1981biometry}
{\sc Sokal, R., and Rohlf, F.}
\newblock Biometry: Principles and practise of statistics in biological
  research wh freeman \& co.
\newblock {\em San Francisco,\/} (1981).

\bibitem{Sonnenschein2000}
{\sc Sonnenschein, C., and Soto, A.~M.}
\newblock Somatic mutation theory of carcinogenesis: {Why} it should be dropped
  and replaced.
\newblock {\em Mol. Carcinog. 29}, 4 (2000), 205--211.

\bibitem{bre}
{\sc Speer, J.~F., Petrosky, V.~E., Retsky, M.~W., and Wardwell, R.~H.}
\newblock A stochastic numerical model of breast cancer growth that simulates
  clinical data.
\newblock {\em Cancer Res. 44}, 9 (1984), 4124--4130.

\bibitem{treat}
{\sc Spina, S., Giorno, V., Rom{\'a}n-Rom{\'a}n, P., and Torres-Ruiz, F.}
\newblock A stochastic model of cancer growth subject to an intermittent
  treatment with combined effects: Reduction in tumor size and rise in growth
  rate.
\newblock {\em Bull. Math. Biol. 76}, 11 (2014), 2711--2736.

\bibitem{st1989analysis}
{\sc St, L., Wold, S., et~al.}
\newblock Analysis of variance (anova).
\newblock {\em Chemometrics and intelligent laboratory systems 6}, 4 (1989),
  259--272.

\bibitem{tabassum}
{\sc Tabassum, D.~P., and Polyak, K.}
\newblock Tumorigenesis: {It} takes a village.
\newblock {\em Nat. Rev. Cancer 15}, 8 (2015), 473--483.

\bibitem{model}
{\sc Yorke, E.~D., Fuks, Z., Norton, L., Whitmore, W., and Ling, C.~C.}
\newblock Modeling the development of metastases from primary and locally
  recurrent tumors: {Comparison} with a clinical data base for prostatic
  cancer.
\newblock {\em Cancer Res. 53}, 13 (1993), 2987--2993.

\bibitem{Zheng}
{\sc Zheng, Q.}
\newblock Progress of a half century in the study of the
  {L}uria--{D}elbr{\"u}ck distribution.
\newblock {\em Math. Biosci. 162}, 1 (1999), 1--32.

\bibitem{zhou2014multi}
{\sc Zhou, D., Wang, Y., and Wu, B.}
\newblock A multi-phenotypic cancer model with cell plasticity.
\newblock {\em J. Theor. Biol. 357\/} (2014), 35--45.

\bibitem{zhou2014nonequilibrium}
{\sc Zhou, J.~X., Pisco, A.~O., Qian, H., and Huang, S.}
\newblock Nonequilibrium population dynamics of phenotype conversion of cancer
  cells.
\newblock {\em PLOS ONE 9}, 12 (2014), e110714.

\end{thebibliography}

\end{document}